\documentclass[3p]{article}
\usepackage[T1]{fontenc}

\usepackage[a4paper,top=2cm,bottom=2cm,left=3cm,right=3cm,marginparwidth=1.75cm]{geometry}
 
\usepackage{amsmath}
\usepackage{graphicx}
\usepackage{hyperref}
\usepackage{float} 
\usepackage{easyReview}
\usepackage{amssymb}
\usepackage{booktabs}
\usepackage{subcaption}
\usepackage{caption}
\usepackage[affil-it]{authblk}

\usepackage{apacite}
\usepackage{natbib}

\title{\bfseries \normalsize Access graph: a novel graph representation of public transport networks for accessibility analysis}

\author[1]{Tina Šfiligoj*}
\author[2]{Aljoša Peperko}
\author[3]{Oded Cats}

\affil[1]{Faculty of Maritime Studies and Transport, University of Ljubljana, Slovenia}
\affil[2]{Faculty of Mechanical Engineering, University of Ljubljana and Institute of Mathematics, Physics and Mechanics, Slovenia}
\affil[3]{Department of Transport \& Planning, Delft University of Technology, The Netherlands}

\date{\vspace{-5ex}}

\begin{document}
\maketitle

\section*{Abstract}\small

Accessibility, defined as travel impedance between spatially dispersed opportunities for activity, is one of the main determinants of public transport use. In-depth understanding of its properties is crucial for optimal public transport systems planning and design. Although the concept has been around for decades and there is a large body of literature on accessibility operationalisation and measurement, a unified approach is lacking. To this end, we introduce a novel graph representation of public transport networks, termed the Access Graph, or A-space, based on the generalised travel times between nodes. We introduce an edge between two nodes in the access graph if the travel time between them is below a certain threshold time budget. In this representation, node degree directly measures the number of nodes reachable within a predetermined time, reproducing the cumulative opportunities measure of access at each specific value of the time budget. We study the threshold-dependent degree distribution of the access graph, focusing on the average degree and the changes in distributions between consecutive time steps. We define a set of accessibility indicators, as well as access equity indicators. The indicators are observed at two characteristic times; the first is based on the evolution of access graph topology and pertaining to the point of degree saturation, reflecting system performance, and the second from the passengers' perspective. We apply the methodology to a dataset of 51 metro networks worldwide. The new representation addresses accessibility at the network structure level, offering a conceptual framework for unified accessibility studies.

\section*{Highlights}

\begin{itemize}
    \item We introduce a novel graph representation of PTNs, the Access graph, or A-space.
    \item The A-space formulation addresses access at the level of the network structure.
    \item Provides an algorithmic approach to accessibility with interpretable indicators.
    \item Presents a unified framework for accessibility analysis rooted in network science.
    \item Applies the methodology on a data set of 51 metro networks worldwide.
\end{itemize}





\textbf{Keywords}: public transport, accessibility, network science, complex systems, metro network, equity

\section{Introduction}

Accessibility is key in providing individuals with the opportunity to take part in employment, education and recreational activities. The key objective of transport systems and public transport services in particular is therefore defined in terms of providing access. The Hansen definition of accessibility as travel impedance between spatially dispersed opportunities for activity is at the core of defining accessibility indicators \citep{hansen1959accessibility}. It has been consistently in use since its conception as evident from a large body of literature, and has facilitated accessibility-based planning \citep{geurs2004accessibility}. In Hansen-like measures of access, also termed primal access, travel time is the independent variable and number of reachable opportunities the dependent variable. Complementary to this, dual access measures invert the independent and dependent variables and measures the time needed to reach a predetermined number of nodes \citep{cui2020primal}.


Despite its wide theoretical and empirical developments, the exact measurement of accessibility remains an open problem. Although the general idea of accessibility is straightforward, the concrete operationalisation and related indicators vary significantly between implementations. The exact definitions and functions of the travel impedance differ, while determining reliable values for the number of opportunities is even more difficult to perform in a unified manner \citep{lahoorpoor2022access}.

This has led to recent efforts to establish unified approaches to accessibility analysis. \citep{levinson2020towards} proposed a general model across different modalities and categories of activity. This was followed by proposing a method which allows for the comparison of different modes of urban accessibility across different cities \citep{wu2021urban}. In \citep{soukhov2025family}, the authors return to the spatial principles of accessibility and review metrics within a physical framework, highlighting the need for scientifically and mathematically rigorous approaches to access measurement. In \citep{van2016accessible, van2022accessibility}, defining unified accessibility indicators has been identified as a major research direction in the coming decades. The lack of unified concepts and measures is seen as a key obstacle toward accessibility-based planning in practice (e.g. \citep{boisjoly2017insider}, \citep{handy2020accessibility}). 
In the afore-referenced works, the overarching thread is the recognition of the importance of the conceptual framework underlying accessibility studies - first its understanding, then its measurement.

Accessibility is fundamentally about connectedness between spatially dispersed locations, which is naturally modelled with graphs. We argue that a network science-based framework will offer new possibilities to conceptualise access by providing a framework that is both deeply rooted in physics, and able to model and predict social and technical phenomena.

The preceding discussion on challenges in accessibility research and planning is especially relevant in the context of public transport network (PTN) planning. Public transport offers a sustainable and space-efficient means of travel, and it is paramount to design accessible PTNs for sustainable and livable cities.
Graph-theoretical and network science approaches have become increasingly important in the field of public transport systems research. Although network science methods have been present in PTN research for several decades and gained traction since the publication of a seminal paper by \cite{von2009public}, there has been a widely recognised concern that this research has been driven primarily by network scientists using PTNs as real-world examples for theoretical developments (e.g. \citep{dupuy2013network}). As a result, the practical implications for transportation researchers and planners have often been overlooked.

Since the 2010s, this gap has begun closing down with transport researchers increasingly using network science approaches in modelling and studying PTN properties \citep{derrible2011applications, ding2019application}. This is especially evident in network robustness and vulnerability studies \citep{adjetey2016model, cats2016robustness, cats2017robustness, cats2020metropolitan}. In contrast, network-based accessibility research has received comparatively less attention. Past studies focused either on analyzing the relative importance of nodes using centrality indicators \citep{cats2017topological, vsfiligoj2025node} or compared different networks using an aggregate metric such as the average shortest path from each node to all other nodes, known as network (in)efficiency \citep{dimitrov2016method, de2019public}. In a pioneering study, \cite{luo2019integrating} suggested connecting network science and accessibility by calculating average shortest paths for each node to all other nodes, providing a Hansen-like indicator similar to closeness centrality. Their application of the method to eight tram networks worldwide demonstrated the potential of network-based methods for comparative assessment. An alternative approach to accessibility that takes into account different components of travel time was presented in \citep{kujala2018travel} where the authors investigated accessibility in terms of Pareto-optimal journeys on temporal networks \citep{kujala2018travel}. As evident from recent literature, the increasing availability of standardised timetable data in the General Transit Feed Specification (GTFS) format has made accessibility studies increasingly accessible \citep{GoogleGTFS}.


In this study, we advance the argument for using graph-theoretical approaches for accessibility research. While the development and subsequent analysis of the standard PTN graph representations offer original insights into PTN properties, we argue that the existing graph representations of PTNs may not offer the level of detail needed for accessibility analysis. At stop level, the two standard PTN representations are the so-called L- and P-space representations, where nodes represent service stops.
In L-space, or space-of-infrastructure, there is an edge between two nodes if they represent consecutive stops on a line, whereas in P-space, or space-of-service, there is an edge between each pair of nodes lying along the same line (i.e., each line is a clique), reflecting the ability of passengers to travel between nodes without transfer. Edge weights may represent in-vehicle and waiting time, for L-space and P-space, respectively \citep{luo2020can}. However, from the accessibility perspective, where travel impedance should include all components of travel time, it is not straightforward to reconcile those in a single representation \citep{kujala2018travel}.

Related to the previous observation, the network properties used to quanitfy accessibility
do not provide a sufficiently complete and insightful information when comparing accessibility across networks. At the stop level, centrality measures, even those in which the global properties of the network are implicit in their definition, offer a local and isolated node-level metrics. In addition, centrality measures originate from social network analysis, and while they have been used successfully in PT studies, their interpretation in the accessibility context may be obscured to some extent as they cannot be unequivocally translated into meaningful access properties. This may consequently limit the generalisation of these measures to potential new models, such as including land use, and thereby diminish their interpretability. At the network level, average shortest path length and network efficiency (defined as the inverse of the sum of all shortest path lengths) are often understood as global measures of accessibility. However, the results will differ based on whether they are calculated in L- or P-space, and in neither representation can the total travel time of passengers be fully captured.

PTNs are spatial networks \cite{barthelemy2011spatial} therefore their infrastructure obeys constraints of planarity, with a possible exception of an occasional tunnel or viaduct. In planar networks, two edges intersect if and only if the intersection constitutes a node. This is encoded in the L-space representation. However, the operational system overlaid on the infrastructure can have a highly non-planar structure, as is the case in the P-space representation. In the latter, the edges are spatially abstracted, and the P-space representation has consistently been shown to exhibit great predictive powers on passenger flows, offering great relevance to transport researchers and planners (e.g. \citep{luo2020can, vsfiligoj2025node}). In P-space, the node degree directly represents the number of stops reachable without transfer and the path length effectively counts the number of transfers. These properties, crucial for understanding passenger behaviour, cannot be modelled in the infrastructural representation.

In line with the above reasoning, and to address the discussed shortcomings, we propose a novel graph representation of PT networks dedicated for accessibility analysis,
called the \textit{Access Graph}, or \textit{A-space}, where two nodes share an edge if they can be reached from one another within a given travel time budget. The construction of the access graph is based on the matrix of generalised travel obtained from L- and P-space representations. The generalised travel time incorporates three components: in-vehicle time, waiting time, and transfer penalties, the latter two weighted with empirically obtained perceived relative costs.

The resulting representation is time-budget dependent, and the edge set of the access graph will differ for each value of $t_b$.
When time budget equals zero, the access graph will be empty, and when it reaches the maximum generalised travel time, the access graph will form a clique. 
The edges of the access graph can be seen as representing functional connections between nodes.
Importantly, in the resulting graph representation, while retaining the same set of nodes, the edges directly indicate accessibility between two nodes. Specifically, node degree in the access graph directly provides the number of reachable nodes within a specific time budget limit, corresponding to the traditional cumulative opportunities accessibility measure \cite{levinson2020towards}. Based on this representation, we study the temporal dependence of node degree distributions.

Furthermore, the distributional aspect of accessibility is one of the most pressing issues in accessibility-based design, where large (spatial) disparities in access to PT impact available opportunities for different social groups \citep{neutens2015accessibility, el2016cost, karner2018assessing, bittencourt2023evaluating, ponkanen2025spatial}. To this end, the Gini coefficient of the degree distribution is used as an indicator of access inequality \citep{gori2020equity}.

The contribution of this study is threefold:

\begin{enumerate}
    \item \textbf{Methodological:} We introduce a novel PTN graph representation, the \textit{access graph}, or the \textit{$A$-space}, which connects pairs of nodes based on their reachability on generalised time-weighted shortest paths. This is the principal contribution of this work.
    \item \textbf{Applied:} Based on the topology of the access graph, we define a set of accessibility indicators, including access inequality indicators.
    \item \textbf{Empirical:} We apply the methodology to a dataset of 51 metro networks worldwide to study the properties of the access graph and compare accessibility measures across networks.
\end{enumerate}

The remainder of this article is structured as follows. In Section \ref{sec:methodology}, the methodology of construction of the access graph and definitions of the proposed indicators are detailed. In Section \ref{sec:results}, the results of the empirical analysis on the metro dataset are presented. Section \ref{sec:conclusion} concludes this study and offers directions for further research.

\section{Methodology}
\label{sec:methodology}

\subsection{Preliminaries}
Many real-world systems, among them (public) transport systems, can be modelled as networks, comprising of sets of elements and their interconnections. A network is modelled as a graph that possesses certain properties of the real-world objects represented by nodes and edges. A graph is a structure $G(V,E)$ with a set of vertices, or nodes, $V = \{v_1,...,v_N\}$ and a set of edges $E = \{e_{ij}\}$, where an edge $e_{ij} = (v_i, v_j)$ connects nodes $v_i$ and $v_j$. $N = |V|$ and $M = |E|$ are the dimension and the size of the graph, respectively. A graph with no edges is an empty graph, and a graph where all pairs of nodes share an edge, is a complete graph, or a clique. A graph is most commonly represented in matrix form with the adjacency matrix $A$ of dimension $N\times N$ and with entries $a_{ij} = 1$ if there is an edge between nodes $i$ and $j$, and $a_{ij} = 0$ otherwise. A weighted representation provides a generalisation where edges are assigned weights, corresponding to their properties (e.g. cost or strength of connection). Node degree $k(i)$ of node $i$ counts the number of the node's direct neighbours, and equals the sum of the corresponding row in the adjacency matrix, $k(i) = \sum_{j=1}^N a_{ij}$. The $N\times N$ distance matrix of the graph$\mathcal{D}$ contains as elements $d_{ij}$ the lengths of the shortest paths between all pairs of nodes. If the graph is disconnected, the respective elements are set to infinity by convention.



\subsection{PTN graph representations}

There exist four standard graph representations of PTNs: at the node-level, the L- and P-space representations; the C-space representation on route-level; and bipartite B-space representation at the combined stop-route level. The L- and P-space representations were introduced in \citep{sienkiewicz2005statistical} and their properties were analysed for 22 Polish cities. In the seminal work of \cite{von2009public} the authors added the C- and B-space representations and consolidated the use of the four representations in the future PTN network studies that followed. In the following, we briefly describe the four representations and their properties.

\begin{itemize}
    \item \textbf{L-space}: this is the most basic representation in which the nodes represent stops and the links connects stops that are adjacent on a route. It is also termed space-of-infrastructure \cite{luo2019integrating}, reflecting the infrastructural nature of connections.
    \item \textbf{P-space}: similar as in L-space, the nodes represent stops, and there is an edge between each pair of nodes that lie on the same route, i.e., each route forms a clique. Also called space-of-service \cite{luo2019integrating}, it reflects the ability of passengers to travel without transfer, and path length in P-space effectively counts the number of transfers.
    \item \textbf{C-space}: nodes represent routes and there is an edge between two nodes if they share at least one stop. Focuses on route-level behaviour and is suitable for studying synchronisation etc.
    \item \textbf{B-space}: a bibartite representation where the two sets of nodes represent stops and routes, respectively, and there is an edge between two nodes if the corresponding stop lies on the corresponding route. Explicitly connects the two levels.
\end{itemize}

\begin{figure}[H]
	\begin{center}
		\texttt{}\includegraphics[scale=0.6]{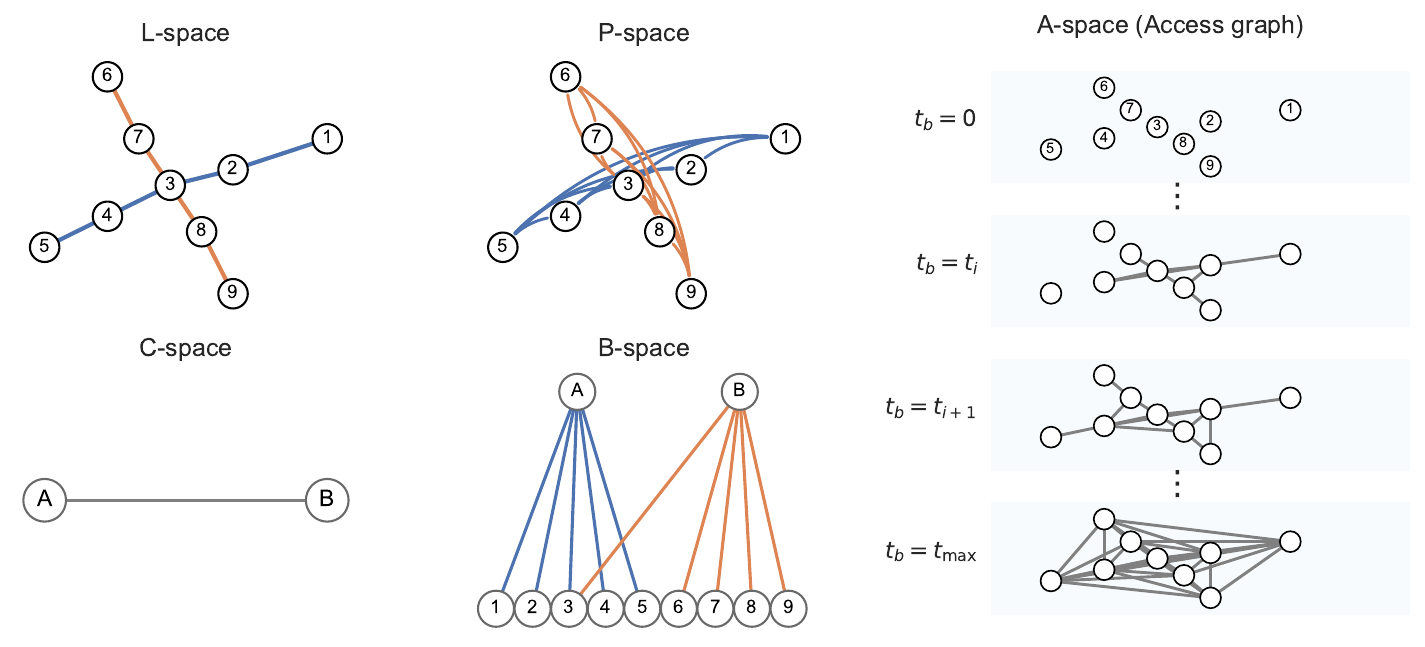}
	\end{center}
	\caption{Overview of existing PTN graph representations and placement of our new A-space representation in the group.}
	\label{fig:spaces}
\end{figure}

The representations can be seen as increasing in levels of spatial abstraction of edges and nodes. In the P-space, the edges reflect an important feature of connectivity and passenger journeys in PT systems, i.e., the transfer behaviour. In C-space, the routes are spatially abstracted into nodes. Each of the representations can offer a different perspective on PTN behaviour. In this vein, we see the access graph, or the A-space representation, as an addition to this family of representations. The node set represents stops, and the edges represent indicate pairs of nodes that can be reached from one another in a given cut-off value of generalised travel time, comprised of in-vehicle and waiting time, plus a time-equivalent transfer penalty. In this sense, the edges in the access graph amount to a form of functional relationships between them the nodes, by design dedicated to accessibility analysis. In line with the discussion above, the edges are spatially abstracted to directly reflect reachability between nodes. In contrast to the existing representations, the access graph is time budget-dependent, resulting in a layered representation where consecutive layers represent the functional connections at consecutive values of cut-off times. The representations and their properties are summarised in Table \ref{tab:representations} and visualised in Figure \ref{fig:spaces}.

\begin{table}[H]
\footnotesize
    \centering
    \begin{tabular}{llll}
        \textbf{Representation} & \textbf{Name} & \textbf{Nodes}  & \textbf{Edges} \\
        \hline
        L-space & Space-of-infrastructure & Stops & Infrastructural connections   \\
        P-space & Space-of-service & Stops & Operational connections   \\
        C-space & Space-of-inter-lining & Routes & Shared stops   \\
        B-space & Space-of-line-interchange & Stops and Routes & Stop lies on route   \\
        \textbf{A-space} & \textbf{Access graph} & Stops & Stops reachable within threshold time   \\
    \end{tabular}
    \caption{A table of PTN graph representations.}
    \label{tab:representations}
\end{table}

In the following section, we describe the construction of the access graph in detail.

\subsection{Access graph}

In PT access studies, generalised travel time is typically comprised of in-vehicle and waiting times, as well as the time-equivalent representation of transfer penalties. In the access graph $\mathcal{G}_A$, the nodes represent the stops of a PTN, and the edge set is obtained depending on the generalised travel times as follows. First, the generalised travel time matrix $\mathcal{D}$ is constructed:

    \begin{enumerate}
        \item Construct the standard graph representations: frequency-weighted P-space representation with edge weights representing service frequencies $f_{ij}$ (given as number of rides per hour) and in-vehicle time-weighted L-space representation with edge weights $t^{L}_{ij}$.
        \item Transform the edge weights of the P-space representation into average waiting time $t^{wait}$ in minutes, defined as half of the headway: $t^{wait}_{ij} = 30/f_{ij}$.
        \item Assign in-vehicle time weights $t^{in-veh}_{ij}$ to P-space edges by calculating the weight of the corresponding path in L-space (summing in-vehicle times on all consecutive links).
        \item Calculate the distance matrix $\mathcal{D}^{P^{wait}}$ of the waiting time-weighted P-space representation.
        \item Determine the distance matrix of in-vehicle times $\mathcal{D}^{P^{in-veh}}$ by calculating the in-vehicle time weights of the paths obtained in the previous step.
        \item Similarly, determine the distance matrix $\mathcal{D}^{P^{u}}$ of corresponding path lengths in unweighted P-space. Note that $([\mathcal{D}^{P^{u}}]_{ij} - 1)$ counts the number of transfers along the path between $i$ and $j$. 
        \item Construct the generalised travel time matrix $\mathcal{D}$: 
        \begin{equation}
            \mathcal{D} = \mathcal{D}_{L^{in-veh}} + w^{\mathrm{wait}} \mathcal{D}_{P^{wait}} + w^{\mathrm{transfer}} (\mathcal{D}_{P^{u}}-J_N),
        \end{equation}
        where $J_N$ is an $N\times N$ matrix of ones, $w^{\mathrm{wait}} = 2$ is the waiting time weight and $w^{\mathrm{transfer}}=5$ min is the transfer penalty. The values are determined based on the literature on empirical behavioral analysis and reflect average passengers' valuation of time \citep{yap2024passenger}.
    \end{enumerate}

Note that the entry $d_{ij} = [\mathcal{D}]_{ij}$ of the generalised travel time matrix represents the generalised travel time between stops $i$ and $j$. In the next step, the time budget $t_b$ is introduced, i.e. the maximum total travel time allowed. In the access graph of $t_b$, there exists an edge between nodes $i$ and $j$ if and only if $d_{ij} \le t_b$. The adjacency matrix $A$ with entries $a_{ij}$ of the access graph is thus obtained from the generalised travel time matrix:

\begin{equation}
    a_{ij} = \begin{cases}
1 & d_{ij} \le t_b\\
0 &\text{otherwise.}
\end{cases}
\end{equation}

The edge set of the access graph is thus time-budget dependent. The $\mathcal{G}_A(t_b)$ obtained in this process is an undirected, unweighted graph. Edges in the access graph connect nodes reachable within the time budget by following the shortest path. Therefore, the node degree $D$ of $\mathcal{G}_A$ reproduces the cumulative opportunities measure of accessibility \cite{el2006access, kapatsila2023resolving}. The access graph as a whole can be thought of as a temporal network with consecutive layers representing the status of the access graph at consecutive values of $t_b$ and thus levels of connectedness at each time step \citep{holme2012temporal}.


\subsection{Access indicators}
\label{sec:indicators}

The evolution of the access graph is observed with increasing the time budget $t_b$ from $0$ to $t_{max}=\max_{i,j} d_{ij}$. At $t_b = 0$ $\mathcal{G}_A$ is an empty graph, and at $t_b = t_{max}$, $\mathcal{G}_A$ is a complete graph. The average degree $\langle k\rangle$ in these boundary cases is $0$ and $N-1$, respectively. To examine global accessibility indicators, we study the change in degree distribution with increasing $t_b$ in the interval $[0, t_{max}]$ for two complementary notions: the growth of average degree and the overall structural change of the network. Based on the shapes of the examined relationships, we identify characteristic values of $t_b$ that mark notable transitions in access behaviour.



The average degree $\langle k\rangle(t_b)$ in the access graph is an increasing function of $t_b$ and by design of real-world transport systems - where the spatial density of stops and routes is highest in a central area and then the distance between adjacent stops as well as parallel routes increases towards the periphery - the edges will typically emerge first in the core and then the periphery gradually becomes better connected. This will result in a sigmoid (or $S$-shape) dependence of $\langle k\rangle$ w.r.t. $t_b$, meaning that at some point there is a point of inflection where the curve changes from convex to concave, and this is where typically the second derivative $\langle k\rangle^{\prime\prime}$ equals zero. This corresponds to the maximum of the first derivative $\langle k\rangle^{\prime} = \frac{d\langle k\rangle}{dt_b}$, and we will take the value of $t_b$ where this maximum occurs as an access indicator. Note that here we are working with discrete time steps and the derivatives are approximated with difference quotients $\frac{\Delta \langle k\rangle}{\Delta t_b}$. From the access perspective, the point where the maximum of the first derivative is achieved is interesting. This point can be understood as the time where the network has reached an "accessible state" through a second-order phase transition, after which the connections between the remaining pairs of nodes emerge at a (distinctively) lower rate. This perspective gives insights into accessibility behaviour from the system performance point of view.

For a complementary perspective, we adopt the passengers' point of view by considering characteristic values of $t_b$ which correspond to typical commute times. Specifically, we observe the properties of the access graph at $t_b=30$ min. The reasoning behind this choice is that the individual daily time budget does not increase, or increases only marginally with city size, and was found to be around one hour \citep{ahmed2014seventy}. Thresholds near the average commuting times were set as cut-off values for the cumulative opportunities measures of access in past studies, for example in \citep{kapatsila2023resolving}.

Based on this reasoning, we propose a set of global accessibility indicators. First, the value of $t_b$ at the occurrence of the point of maximum average degree growth, $t_M$, is proposed as an absolute-value dual access indicator. The reasoning behind taking an absolute value of time for performance measurement is again based on the 30 minute goal regardless of city (or network) size.
For a complementary measure, an additional indicator $\delta t_M$ is proposed that measures the point of maximum growth relative to the maximum generalised travel time observed in the network, i.e.  $\delta t_M^{\Delta} = t_M^{\Delta}/t_{max}$. This gives a measure of performance that is normalised in relation to the farthest away origin-destination pair (which is equivalent to the notion of graph diameter), i.e. how early in the interval the maximal access growth is reached.
Note that lower values of $t_M$ and $\delta t_M$ generally imply higher accessibility. In addition to the values of the (relative) time budgets at maximal growth, the values of the average degree at $t_M$ and $t_b=30$ min, $\langle k\rangle_M$ and $\langle k\rangle_{30}$, respectively, constitute complementary primal indicators.
We emphasise that although the indicators are based on the average degree, the choice of the values at which we measure the average degree is based on the overall evolution of the access graph. In this sense, the indicators capture the global behaviour of accessibility within this averaged measure.

Alongside the average degree, we examine the $t_b$-dependent degree distributions of the access graph for the purpose of access equity assessment. A standard measure of inequality is the Gini coefficient, defined as:

\begin{equation}
    G = \frac{\sum_{i=1}^n\sum_{j=1}^n |x_i - x_j|}{2n^2\overline{x}},
\end{equation}
where $\overline{x} = \frac{1}{n}\sum_{i=1}^n x_i$ is the distribution mean \citep{gori2020equity}. $G$ can assume values in the interval $[0, 1]$, where $G=0$ represents perfect equity and $G=1$ means maximum disparity. In this study, the values of the Gini coefficients of the node degree distributions of the access graph are observed. Following a similar reasoning as above, $G_M$ and $G_{30}$, representing the values of the Gini coefficients of the degree distribution of the access graph at $t_b = t_M$ and $t_b = 30$ min, respectively, are proposed as indicators of access inequality. The Gini coefficient is chosen here rather than skewness, because the degree distributions in our empirical analysis were found to often exhibit a bimodal shape.

In addition, we investigate how degree distributions evolve for consecutive time steps. We adopt the Jensen-Shannon divergence (JSD) as a metric of statistical distance between the distributions, which is commonly used as a measure to quantify structural differences across layers in multiplex networks (see e.g. \citep{de2015structural}). It is defined as:
\begin{equation}
\label{eq:jsd}
\operatorname{JSD}(P\|Q)
= \frac{1}{2}\,\operatorname{KL}\!\left(P\middle\|M\right)
+ \frac{1}{2}\,\operatorname{KL}\!\left(Q\middle\|M\right),
\qquad
M = \frac{1}{2}(P+Q),
\end{equation}
where $\operatorname{KL}$ is the Kullback-Leibler divergence, which measures relative entropy of two probability distribuitions $P$ and $Q$. For a discrete distribution it is defined as:
\begin{equation}
\label{eq:kl_discrete}
\operatorname{KL}(P\|Q)
= \sum_{i} P(i)\,\log_2\!\frac{P(i)}{Q(i)}.
\end{equation}
The JSD is then the symmetrised version of KL.
JSD measures the "surprise value" of distribution $P$ given the distribution $Q$ based on the relative entropy and is thus sensitive to the overall degree distribution shape. For the access graph, $Q(i)$ and $P(i)$ will be the degree distributions at consecutive values of $t_b$: if $Q(i)$ is the degree distribution at $t_b=t_j$ (i.e. $j$th step) then $P(i)$ is the degree distribution at $t_b=t_{j+1}$. The values of JSD range from 0 when the two distributions are identical, to 1 (if the logarithm base is 2, which is typically the case in information entropy measures). The value 1 will only be reached for two distributions at disjoint domains, which does not happen for the degree distributions on the same set of nodes as is the case for the access graph (i.e., the domain is $[0,N-1]$ for all $t_b$ values).

In contrast to the average degree, $JSD$ is not a monotonous function of $t_b$ and is sensitive to the choice of the time step. For this reason, a single value of $t_b$ where JSD reaches a maximum is not a robust measure. To overcome this, we smooth the curve with a sliding window and observe the cumulative JSD in the intervals with the width equal to 10\% of the total $t_b$ interval, i.e., $w=0.1t_{max}$. We then calculate the cumulative values with the sliding window and identify the interval of width $w$ at which the total change in JSD is the largest. We then take the midpoint $t_b$ value of this interval as a dual access indicator $t_{JS}$. Then the time $t_{JS}\pm w/2$ is the time of the fastest change in access graph structure. Similar as in the case of $t_M$, we propose the time relative to the maximum time budget, $\delta t_{JS} = t_{JS}/t_{max}$, and the average degree at $t_{JS}$, $\langle k\rangle_{JS}$, as additional indicators.

The accessibility indicators used in the analysis are summarised in Table \ref{tab:indicators}.

\begin{table}[H]
\centering
\footnotesize
\caption{Accessibility indicators used in the analysis.}
\begin{tabular}{p{1.4cm}p{5cm}p{7cm}}
\hline
\textbf{Notation} & \textbf{Name} & \textbf{Definition} \\ \hline
$t_{max}$ & Network diameter & maximum generalised travel time in the network\\
$t_M$ & Access inflection point & time budget at maximum growth of average degree \\
$\delta t_M$ &Normalised access inflection point& $t_M/t_{max}$; $t_M$ relative to maximum generalised travel time  \\
$t_{JS}$ & Access structural change point & characteristic time budget at maximum structural change in of the access graph \\
$\delta t_{JS}$ & Normalised access structural point&$t_{JS}/t_{max}$; $t_{JS}$ relative to maximum generalised travel time  \\
$\langle k\rangle_M$ &Access at inflection point& average degree of the access graph at $t_M$ \\
$\langle k\rangle_{JS}$ &Access at structural change point & average degree of the access graph at $t_{JS}$ \\
$\langle k\rangle_{30}$ &30-min access& average degree of the access graph at $t_b=30$ min\\
$G_M$ &Inflection access inequality & the Gini coefficient of the degree distribution at $t_M$ \\
$G_{30}$ &30-min access inequality & the Gini coefficient of the degree distribution at $t_b=30$ min 
\end{tabular}
\label{tab:indicators}
\end{table}






\section{Results and analysis}
\label{sec:results}

The proposed methodology is applied to a dataset of metro networks of 51 cities worldwide. We use publicly available in-vehicle time-weighted L-space and frequency-weighted P-space representations which were constructed using GTFS data \citep{lspace,pspace}. For each metro network, access graphs are built for varying time budgets. $t_b$ is varied in the interval $[0, t_{max}]$, where $t_{max}$ is the maximum generalised travel time in the network, with progressively increasing steps of 2 minutes. The Python NetworkX library was used for graph construction and analysis \citep{hagberg2008exploring}. In the following, we first illustrate our novel A-space and the proposed global indicators using a detailed explanation of the results and a comparison between the metro networks of Oslo and Vienna (Section \ref{sec:amsterdam}) before then proceeding to the complete set of results of the empirical analysis for the full metro dataset (Section \ref{sec:all51}).

\subsection{Access graph illustration: case studies of Oslo and Vienna metro networks}
\label{sec:amsterdam}

We illustrate our approach for metro networks in two European cities: Oslo and Vienna. The networks were chosen based on their similar dimension of approximately 100 nodes, which is large enough for the network to exhibit smooth behaviour yet small enough to allow for meaningful visualisations. The Oslo network has $N=101$ nodes and a maximum generalised travel time of $t_{max}=92$ minutes, and for Vienna the values are $N=98$ and $t_{max}=74$ minutes.

\begin{figure}[H]
	\begin{center}
		\texttt{}\includegraphics[scale=0.5]{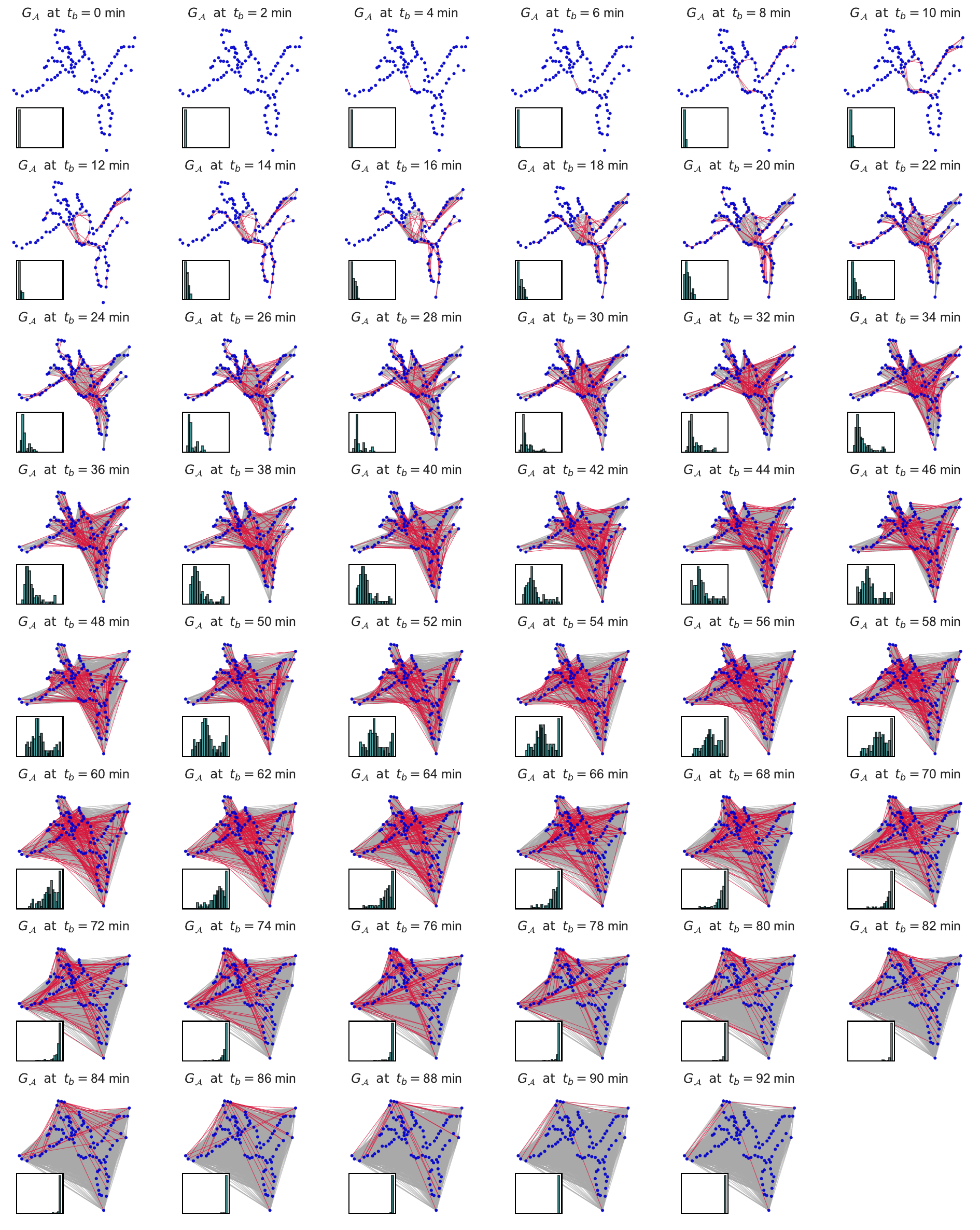}
	\end{center}
	\caption{A visualization of the time budget-dependent access graph for the Oslo metro. In successive subplots, access graphs at respective values of $t_b$, as indicated in each subplot title, are shown. The edges that first appear at a given $t_b$ are shown in red. Other edges (i.e., those present in $\mathcal{G}_A$ already earlier than at $t_b$) are shown in grey. The Oslo metro has $N=101$ nodes and a maximum generalised travel time of $t_{max} = 92$ min.}
	\label{fig:vis_oslo}
\end{figure}

\begin{figure}[H]
	\begin{center}
		\texttt{}\includegraphics[scale=0.4]{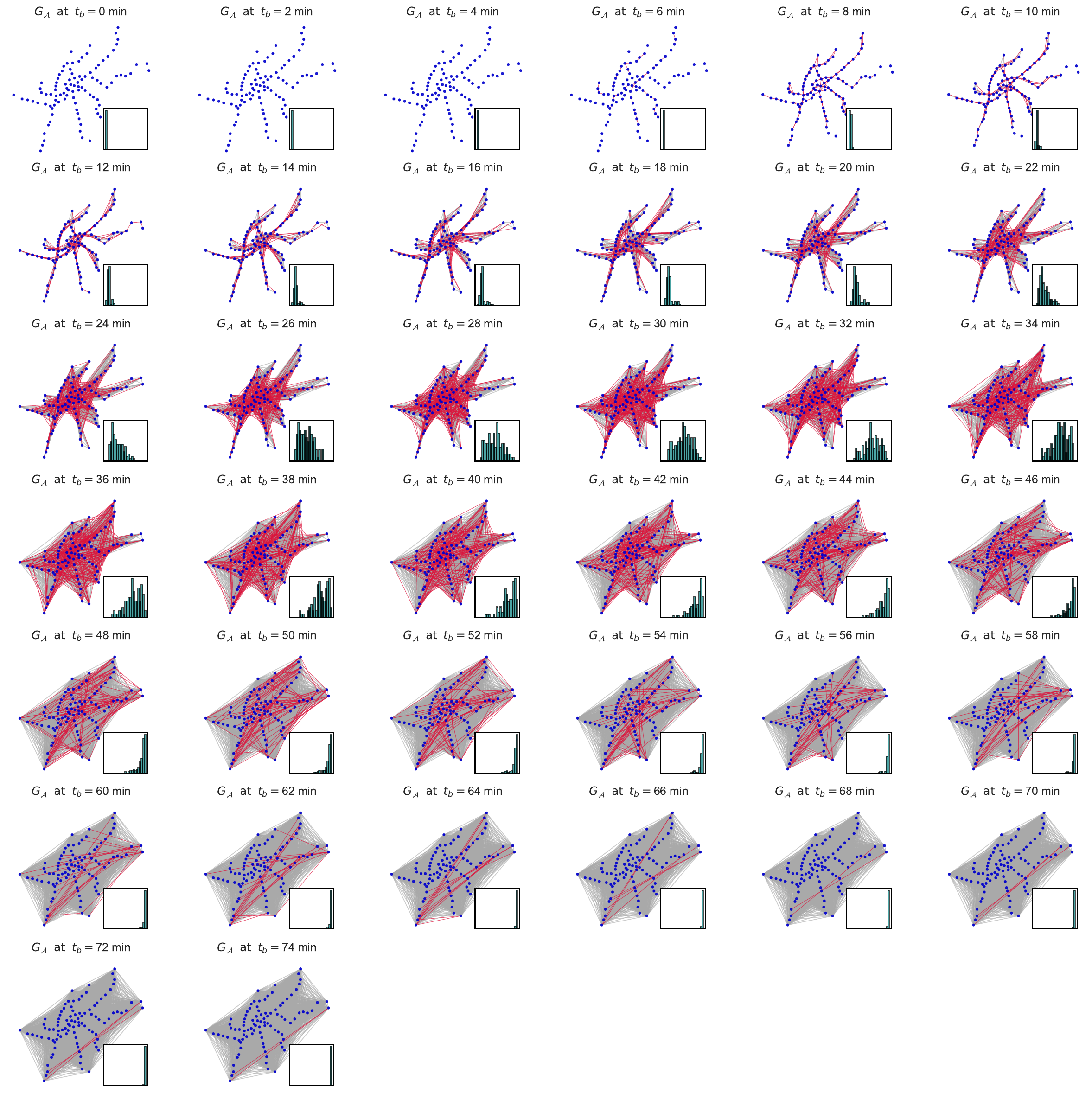}
	\end{center}
	\caption{A visualization of the time budget-dependent access graph for the Vienna metro. 
    In successive subplots, access graphs at respective values of $t_b$, as indicated in each subplot title, are shown. The edges that first appear at any given $t_b$ are shown in red. Other edges (i.e., those present in $\mathcal{G}_A$ already earlier than at $t_b$) are shown in grey. In each subplot, the degree distribution of $\mathcal{G}_A$ is shown in the inset plot. The Vienna metro has $N=98$ nodes and a maximum generalised travel time of $t_{max} = 74$ min.}
	\label{fig:vis_vienna}
\end{figure}

The growth of the access graph in progressive steps of 2 minutes of $t_b$ is shown in Figures \ref{fig:vis_oslo} for Oslo and \ref{fig:vis_vienna} for Vienna. Each subplot shows the access graph $\mathcal{G}_A(t_b)$. The edges that first appear at the respective $t_b$ are shown in red, and the edges that were added earlier are shown in grey. In each subplot, there is an inset plot of the degree distribution at the respective value of $t_b$.

\begin{figure}[H]
	\begin{center}
		\texttt{}\includegraphics[scale=0.85]{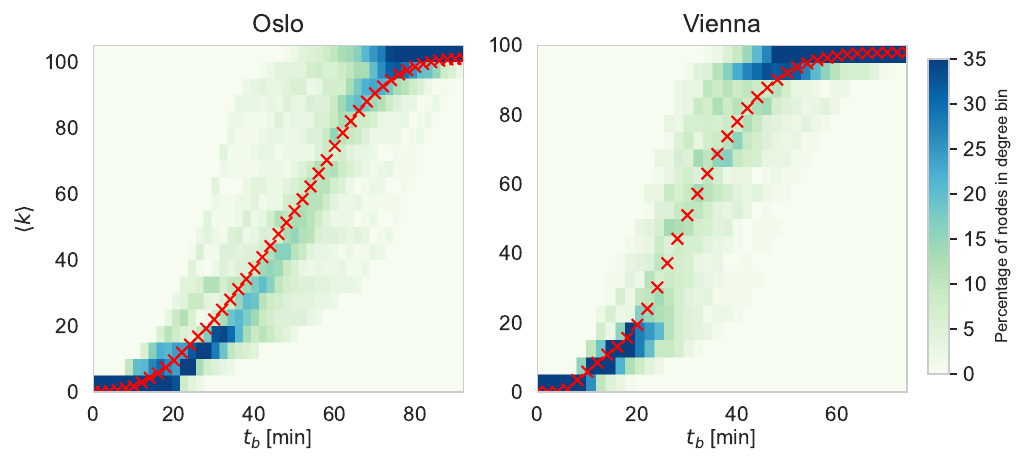}
	\end{center}
	\caption{Heatmaps of degree distributions with varying time budgets for the Oslo (left) and Vienna (right) metros. $t_b$ is increased in steps of 2 minutes. The $x$-axis represents the time budget, and the $y$-axis represents the node degree in the respective access graph. Heatmap cell color represents the percentage of nodes in each degree bin. Thus, each vertical column represents the color-coded histogram of the node degree distribution. The average degree at each $t_b$ bin is plotted with red markers.}
	\label{fig:example_heatmaps}
\end{figure}

This behaviour is more compactly presented as heatmaps in Figure \ref{fig:example_heatmaps}. In the heatmap, the $x$-axis represents the time budget $t_b$, and $y$-axis represents the degree $k$ in the respective access graph. The width of $t_b$ and $k$ bins are 2 minutes and 5, respectively. The colors of the heatmap cells represent the percentage of nodes in each degree bin. The average degree at each $t_b$ is shown with a red marker.


At lower values of $t_b$, the degree distribution is skewed heavily to the right, which is also observed in Figures \ref{fig:vis_oslo} and \ref{fig:vis_vienna}.
As the travel time budget increases, the values of the average degree, along with the minimum and maximum degree, increase and the degree distribution is progressively skewed to the left. The distributions for the intermediate values of $t_b$ are often bimodal or even multimodal with several local maxima, which is observed for Oslo for example in the histogram at $t_b=50$ min, where the distribution is bimodal (one larger peak at $k\approx40$, and a smaller peak at $k\approx 90$). Conversely, the degree distributions for Vienna exhibit roughly unimodal behaviour throughout the interval, where the peak of the distribution moves progressively to higher values. This can also be observed in the heatmap plots where for Oslo, several distinct $S$-shapes can be observed, marking observably faster growth of degree of a group of nodes to the left of the average degree and the bulk of the distribution. For the Vienna metro, the bulk stays around the average degree and gradually decreases away from it on both sides.
We also observe steeper growth of average degree for the Vienna network early in the interval.
From the above, we qualitatively expect that the Vienna metro is more accessible and has higher access equity than the Oslo metro.

\begin{figure}[H]
    \centering

    \begin{subfigure}[t]{0.48\textwidth} 
        \centering
        \includegraphics[width=\textwidth]{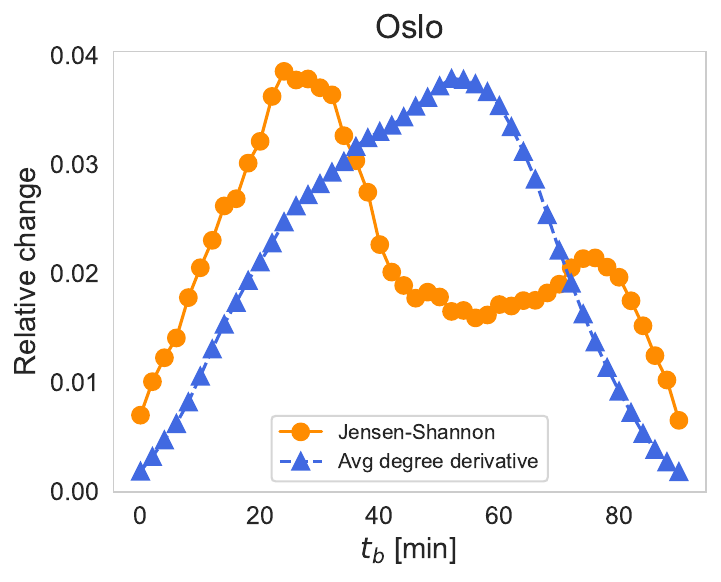}
        \caption{}
        \label{fig:sub1}
    \end{subfigure}
    \begin{subfigure}[t]{0.48\textwidth} 
        \centering
        \includegraphics[width=\textwidth]{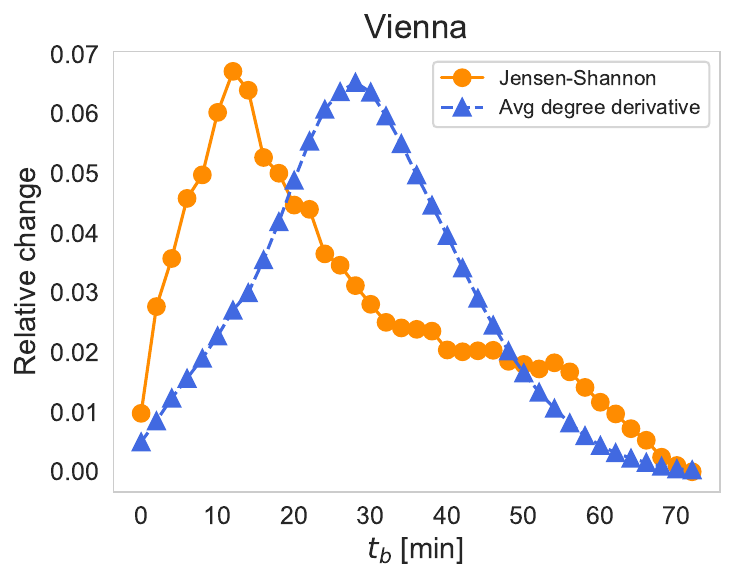}
        \caption{}
        \label{fig:sub2}
    \end{subfigure}
    \caption{Oslo (left) and Vienna (right).}
    \label{fig:oslo_vienna_metrics}
\end{figure}

We quantitatively compare the pace at which access progresses in both network by examining how node degree distributions change across progressive Access graphs and based on the time at maximum growth of average degree, i.e., the point where its first derivative is maximal. To examine the behaviour of the derivative in more detail, we observe the shape of the average degree derivative in the whole $t_b$ interval. For a complementary view, sensitive to the overall structure, we examine the Jensen-Shannon divergence alongside it. The results are shown in Figure \ref{fig:oslo_vienna_metrics}. We see that the shape of $\langle k \rangle^{\prime}(t_b)$ is wider for Oslo compared to Vienna, and the peak for Vienna is more prominent. This is in line with the disparities in level of accessibility for the Oslo network where the group of better connected nodes drives the change faster, but is spread out by the slower growth in degree for the majority of nodes. For the Vienna metro, the behaviour is much more regular, in line with observed higher equity of degree growth across the whole network.

An interesting observation is that JSD offers a qualitatively different perspective. For Oslo, it has two peaks, first at about $t_b=25$ min and the second near $t_b=80$ min. This means that there are two points at which the overall structure of the degree distribution changes significantly. Returning to Figure \ref{fig:vis_oslo}, we see that at $t_b=24$ min, the smaller peak to the right emerges, and by $t_b=76$ min, the smaller peak (now to the left) is almost gone, and after that point, the majority of the nodes have the maximum degree and the distribution is heavily left-skewed, with the remaining nodes catching up in connectedness. For Vienna, the only prominent JSD peak occurs at about $t_b=15$ min. This coincides with the value at which the access graph becomes fully connected. Note that in the access graph, by definition, degree is a monotonically increasing function, so positive JSD here always means an increase in accessibility, and a peak in JSD corresponds to a larger change. 

\subsection{Application: Results for 51 metro networks worldwide}
\label{sec:all51}

\begin{figure}[H]
	\begin{center}
		\texttt{}\includegraphics[scale=0.425]{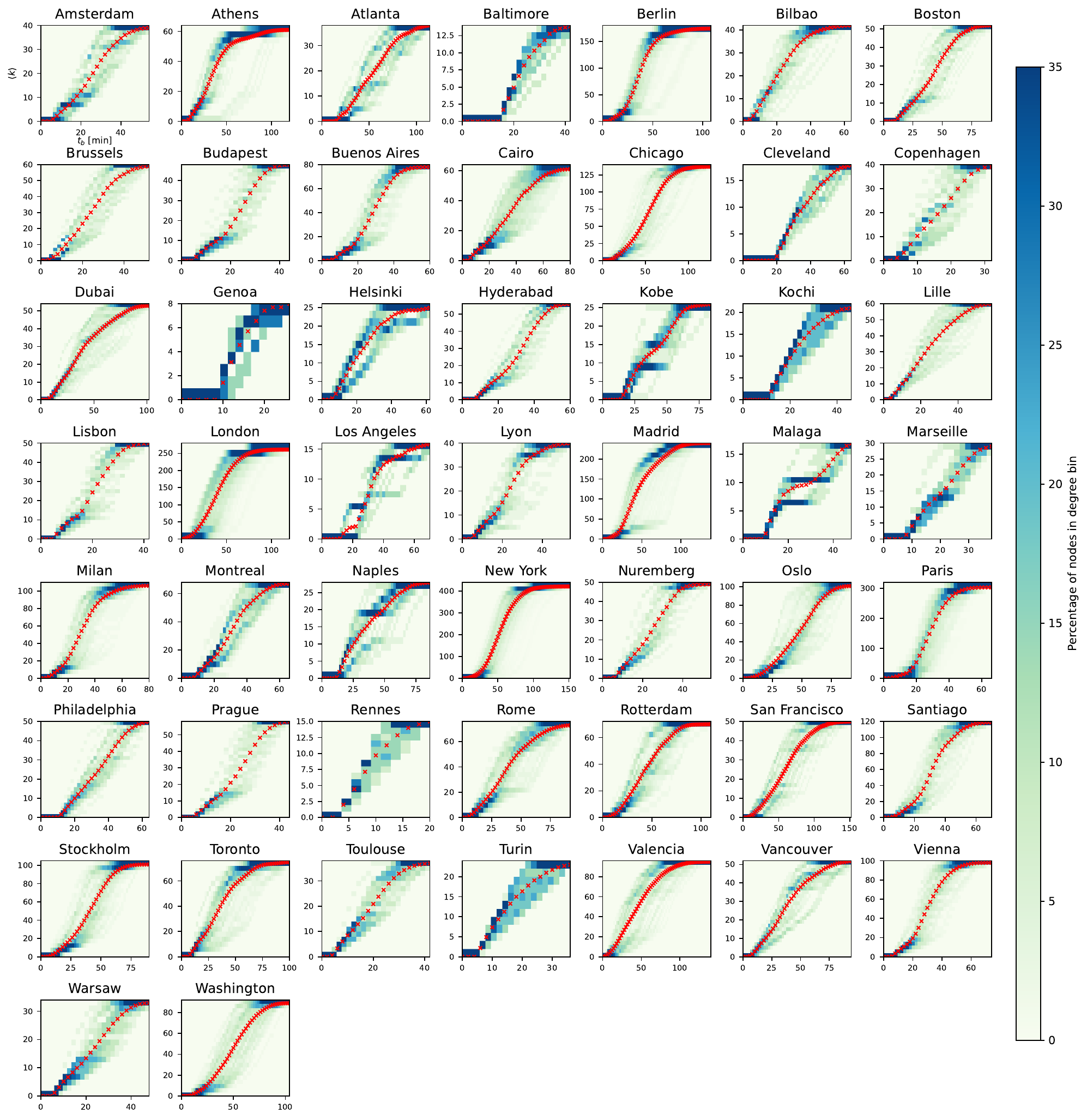}
	\end{center}
	\caption{Degree distributions of access graphs with varying $t_b$. Each subplot shows the time evolution of the access graph with the heatmaps representing node degree distributions. $t_b$ is increased in steps of 2 minutes. In each subplot, the $x$-axis represents the time budget, and $y$-axis represents the degree in the corresponding access graph. Heatmap cell colors represent the percentage of nodes in each degree bin. Average degree of the access graph at $t_b$ is shown with red markers.}
	\label{fig:degdist}
\end{figure}

\begin{figure}[H]
	\begin{center}
		\texttt{}\includegraphics[scale=0.425]{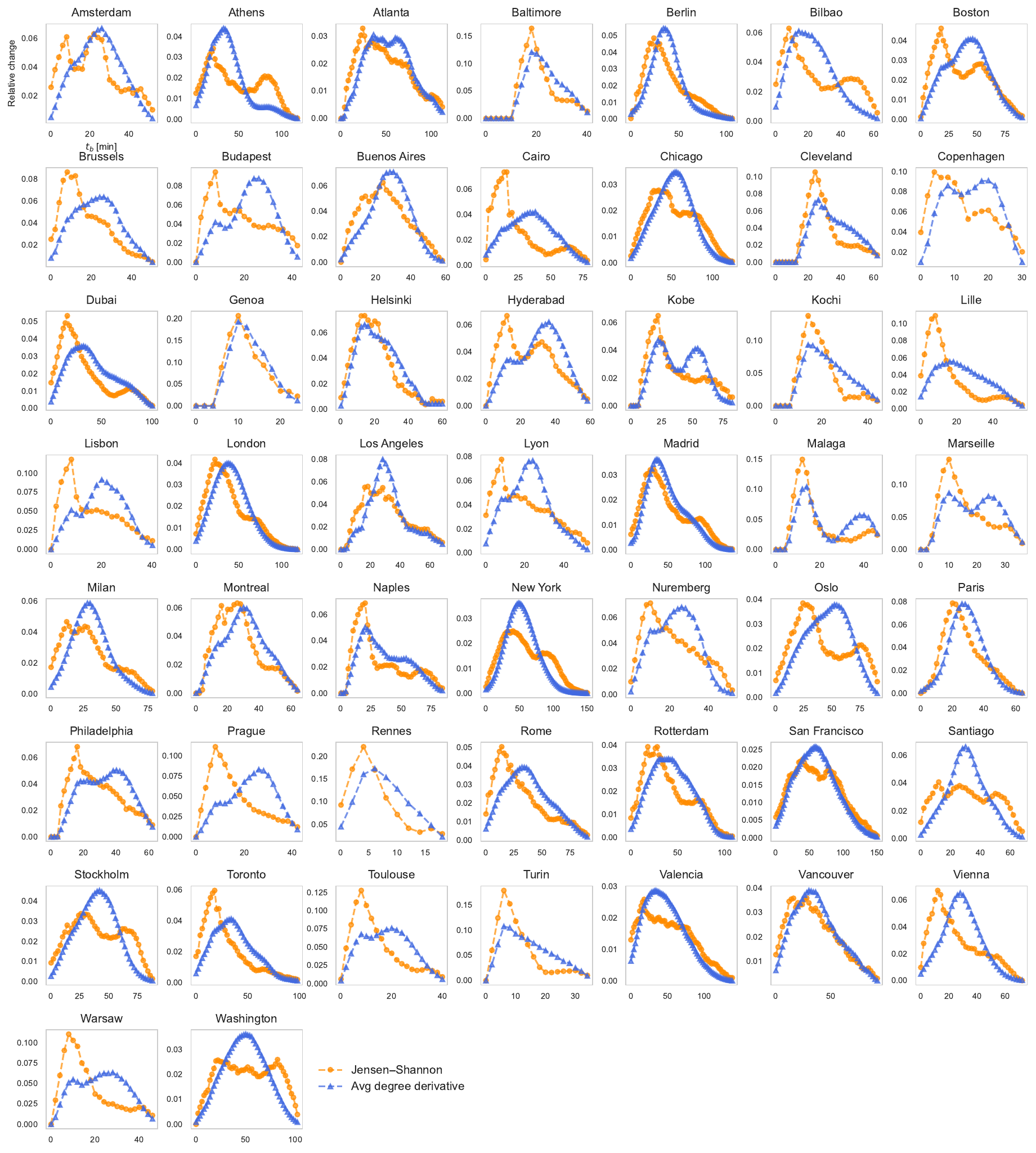}
	\end{center}
	\caption{Average degree derivative and Jensen-Shannon divergence vs. $t_b$ plots for all cities.}
	\label{fig:all_cities_indicators}
\end{figure}

In this section, we turn to numerical results for the access indicators introduced in \ref{sec:indicators} for all 51 metro networks included in our analysis. The discussion in the previous section (\ref{sec:amsterdam}) serves as a basis for interpreting the reported values. The heatmap plots of degree distribution and the $\langle k\rangle^{\prime}(t_b)$ and $JSD(t_b)$ plots that were shown for Oslo and Vienna in Figures \ref{fig:example_heatmaps} and \ref{fig:oslo_vienna_metrics} are given for all cities in Figures \ref{fig:degdist} and \ref{fig:all_cities_indicators}. In the latter figures, each of the subplots represents a separate network (city) and shows the time evolution of the observed quantities. 

\subsubsection*{Average degree-based access indicators}

In almost all cases, a logistic-like $S$-shaped convergence of average degree $\langle k\rangle$ to the maximum degree is observed. 
The most regular $S$-shaped behaviour is observed for the largest networks (New York, London and Paris, among others). In these cases, the value of $t_M$ corresponds to the inflection point (i.e. the "turning point" of the $S$-curve). Several networks exhibit a more complex behaviour with multiple points of inflection in $\langle k\rangle(t_b)$ (e.g. Athens, Los Angeles and Naples), implying several local extrema of $\langle k\rangle(t_b)$. Upon closer inspection, we observe line-level effects (e.g. line length and the importance of transfers in connecting node pairs) and the general topology of the L-space graph as likely predictors of this behaviour.

The numerical values of all indicators are summarised in Table \ref{tab:vars}. The networks are of remarkably diverse dimensions, ranging from $N=8$ (Genoa) to $N=421$ (New York). Maximum generalised travel time ranges from $t_{max}=20$ (Rennes) to $t_{max}=152$ minutes (New York, San Francisco). Large metropolitan areas are expected to have large values of $t_{max}$, e.g. 120 and 138 minutes for London ($N=261$) and Madrid ($N=240$), respectively. An exception is Paris with $N=303$ and $t_{max}=66$ min suggesting a high level of accessibility.
On the other hand, moderately sized networks with high values of $t_{max}$, e.g. Valencia ($N=95, t_{max}=140$), Chicago ($N=137, t_{max}=126$) or Athens ($N=61, t_{max}=120$) are less accessible, with the extreme case of San Francisco with $N=50$ and $t_{max}=152$.

\begin{table}[H]
\centering
\footnotesize
\caption{Values of the studied variables for each network (city). $N$: number of nodes in the network; $t_{max}$: maximum generalised travel time (in minutes); $t_M$: access inflection point; $\delta t_M = \frac{t_M}{t_{max}}$ normalised access inflection point; $t_{JS}$: access structural change point; $\delta t_M = \frac{t_{JS}}{t_{max}}$ normalised access structural point; $\langle k\rangle_M$: access at inflection point; $\langle k\rangle_{JS}$: access at structural change point; $\langle k\rangle_{30}$: 30-min access; $G_M$: inflection access inequality; $G_{30}$: 30-min access inequality.}
\begin{tabular}{lrrrrrrrrrrr}
\toprule
City & $N$ & $t_{max}$ & $t_M$ & $\delta t_M$ & $t_{JS}$ & $\delta t_{JS}$ & $\langle k\rangle_M$ & $\langle k\rangle_{JS}$ & $\langle k\rangle_{30}$ & $G_M$ & $G_{30}$ \\
\midrule
Amsterdam & 39 & 54 & 29 & 0.54 & 7 & 0.13 & 0.67 & 0.06 & 0.67 & 0.18 & 0.18 \\
Athens & 61 & 120 & 31 & 0.26 & 14 & 0.12 & 0.47 & 0.11 & 0.42 & 0.27 & 0.28 \\
Atlanta & 38 & 114 & 39 & 0.34 & 20 & 0.18 & 0.33 & 0.06 & 0.17 & 0.27 & 0.38 \\
Baltimore & 14 & 42 & 21 & 0.50 & 16 & 0.38 & 0.51 & 0.11 & 0.85 & 0.16 & 0.11 \\
Berlin & 174 & 108 & 37 & 0.34 & 19 & 0.18 & 0.55 & 0.10 & 0.31 & 0.22 & 0.29 \\
Bilbao & 42 & 64 & 17 & 0.27 & 5 & 0.08 & 0.42 & 0.02 & 0.76 & 0.20 & 0.13 \\
Boston & 52 & 92 & 45 & 0.49 & 15 & 0.16 & 0.59 & 0.08 & 0.28 & 0.21 & 0.25 \\
Brussels & 59 & 52 & 29 & 0.56 & 11 & 0.21 & 0.72 & 0.17 & 0.72 & 0.16 & 0.16 \\
Budapest & 48 & 44 & 25 & 0.57 & 6 & 0.14 & 0.62 & 0.04 & 0.80 & 0.20 & 0.13 \\
Buenos Aires & 78 & 60 & 29 & 0.48 & 21 & 0.35 & 0.59 & 0.23 & 0.59 & 0.21 & 0.21 \\
Cairo & 61 & 80 & 35 & 0.44 & 12 & 0.15 & 0.55 & 0.12 & 0.42 & 0.22 & 0.24 \\
Chicago & 137 & 126 & 49 & 0.39 & 38 & 0.30 & 0.45 & 0.26 & 0.16 & 0.29 & 0.37 \\
Cleveland & 18 & 64 & 25 & 0.39 & 23 & 0.36 & 0.29 & 0.17 & 0.44 & 0.11 & 0.10 \\
Copenhagen & 39 & 32 & 21 & 0.66 & 4 & 0.12 & 0.79 & 0.03 & 1.00 & 0.14 & 0.00 \\
Dubai & 53 & 102 & 33 & 0.32 & 11 & 0.11 & 0.46 & 0.07 & 0.37 & 0.17 & 0.19 \\
Genoa & 8 & 26 & 11 & 0.42 & 9 & 0.35 & 0.45 & 0.00 & 1.00 & 0.28 & 0.00 \\
Helsinki & 25 & 62 & 11 & 0.18 & 17 & 0.27 & 0.21 & 0.32 & 0.76 & 0.28 & 0.14 \\
Hyderabad & 56 & 60 & 35 & 0.58 & 9 & 0.15 & 0.64 & 0.03 & 0.44 & 0.17 & 0.19 \\
Kobe & 26 & 84 & 21 & 0.25 & 18 & 0.21 & 0.22 & 0.09 & 0.40 & 0.26 & 0.19 \\
Kochi & 21 & 46 & 13 & 0.28 & 12 & 0.26 & 0.20 & 0.09 & 0.82 & 0.12 & 0.11 \\
Lille & 60 & 58 & 15 & 0.26 & 7 & 0.12 & 0.33 & 0.11 & 0.68 & 0.17 & 0.15 \\
Lisbon & 50 & 42 & 19 & 0.45 & 6 & 0.14 & 0.50 & 0.05 & 0.90 & 0.22 & 0.08 \\
London & 261 & 120 & 41 & 0.34 & 16 & 0.13 & 0.56 & 0.09 & 0.31 & 0.24 & 0.37 \\
Los Angeles & 16 & 70 & 31 & 0.44 & 26 & 0.37 & 0.59 & 0.28 & 0.48 & 0.16 & 0.20 \\
Lyon & 40 & 54 & 25 & 0.46 & 7 & 0.13 & 0.63 & 0.09 & 0.78 & 0.20 & 0.14 \\
Madrid & 240 & 138 & 33 & 0.24 & 31 & 0.22 & 0.31 & 0.27 & 0.23 & 0.37 & 0.40 \\
Malaga & 17 & 48 & 11 & 0.23 & 12 & 0.25 & 0.20 & 0.18 & 0.62 & 0.13 & 0.14 \\
Marseille & 29 & 38 & 9 & 0.24 & 10 & 0.26 & 0.15 & 0.14 & 0.90 & 0.17 & 0.08 \\
Milan & 106 & 80 & 29 & 0.36 & 8 & 0.10 & 0.52 & 0.04 & 0.52 & 0.26 & 0.26 \\
Montreal & 67 & 66 & 33 & 0.50 & 13 & 0.20 & 0.62 & 0.08 & 0.49 & 0.17 & 0.21 \\
Naples & 28 & 86 & 17 & 0.20 & 16 & 0.19 & 0.19 & 0.11 & 0.47 & 0.34 & 0.23 \\
New York & 421 & 152 & 51 & 0.34 & 44 & 0.29 & 0.49 & 0.33 & 0.11 & 0.25 & 0.42 \\
Nuremberg & 49 & 54 & 25 & 0.46 & 9 & 0.17 & 0.58 & 0.06 & 0.71 & 0.20 & 0.17 \\
Oslo & 101 & 92 & 59 & 0.64 & 29 & 0.32 & 0.75 & 0.19 & 0.22 & 0.15 & 0.30 \\
Paris & 303 & 66 & 27 & 0.41 & 19 & 0.29 & 0.49 & 0.18 & 0.57 & 0.25 & 0.22 \\
Philadelphia & 50 & 64 & 39 & 0.61 & 13 & 0.20 & 0.65 & 0.03 & 0.42 & 0.14 & 0.18 \\
Prague & 58 & 44 & 27 & 0.61 & 6 & 0.14 & 0.69 & 0.03 & 0.77 & 0.18 & 0.14 \\
Rennes & 15 & 20 & 7 & 0.35 & 3 & 0.15 & 0.51 & 0.12 & 1.00 & 0.12 & 0.00 \\
Rome & 73 & 92 & 35 & 0.38 & 11 & 0.12 & 0.50 & 0.08 & 0.37 & 0.21 & 0.24 \\
Rotterdam & 70 & 110 & 39 & 0.35 & 24 & 0.22 & 0.47 & 0.18 & 0.28 & 0.24 & 0.28 \\
San Francisco & 50 & 152 & 61 & 0.40 & 48 & 0.32 & 0.55 & 0.35 & 0.17 & 0.23 & 0.35 \\
Santiago & 119 & 70 & 31 & 0.44 & 10 & 0.14 & 0.53 & 0.04 & 0.45 & 0.22 & 0.25 \\
Stockholm & 101 & 90 & 45 & 0.50 & 30 & 0.33 & 0.63 & 0.27 & 0.28 & 0.21 & 0.33 \\
Toronto & 75 & 100 & 29 & 0.29 & 13 & 0.13 & 0.41 & 0.09 & 0.41 & 0.28 & 0.28 \\
Toulouse & 37 & 42 & 21 & 0.50 & 6 & 0.14 & 0.66 & 0.09 & 0.91 & 0.16 & 0.07 \\
Turin & 23 & 36 & 7 & 0.19 & 6 & 0.17 & 0.22 & 0.09 & 0.99 & 0.09 & 0.04 \\
Valencia & 95 & 140 & 21 & 0.15 & 11 & 0.08 & 0.20 & 0.07 & 0.32 & 0.40 & 0.33 \\
Vancouver & 52 & 94 & 31 & 0.33 & 25 & 0.27 & 0.42 & 0.26 & 0.38 & 0.20 & 0.20 \\
Vienna & 98 & 74 & 27 & 0.36 & 10 & 0.14 & 0.46 & 0.06 & 0.53 & 0.25 & 0.23 \\
Warsaw & 33 & 48 & 27 & 0.56 & 8 & 0.17 & 0.67 & 0.09 & 0.74 & 0.17 & 0.15 \\
Washington & 89 & 104 & 51 & 0.49 & 27 & 0.26 & 0.56 & 0.16 & 0.19 & 0.25 & 0.35 \\
\bottomrule
\end{tabular}
\label{tab:vars}
\end{table}

Nine of the networks comprise of a single line: these are the metros in Baltimore, Cleveland, Genoa, Helsinki, Kochi, Los Angeles, Malaga, Rennes, and Turin. Their dimension ranges from 8 (Genoa) to 25 (Helsinki). For one-line networks, the growth of the access graph depends only on in-vehicle times as there are no transfers, and waiting time is uniform across the network.

The access inflection point values range from $t_M=7$ (Rennes) to $t_M=61$ minutes (San Francisco). Note that almost half of the networks (23 networks) have $t_M$ in the interval between 25 and 35 minutes, meaning that the operational characteristic time is close to the passengers'-oriented value of 30 minutes. The values of $\delta t_M$ range from $0.15$ (Valencia) $0.66$ (Copenhagen). In general, low values of the normalised access inflection point indicate higher accessibility, but this is not necessarily the case.
For example, Valencia has the lowest value of $\delta t_M$, however the value of access at inflection point $\langle k\rangle_M=0.19$ shows that degree growth is far from saturation. Conversely, the Copenhagen metro reaches the highest value among networks with $\langle k\rangle_M=0.75$ and is one of the best connected networks with $t_{max}=32$ and $t_M=21$ minutes at $N=39$. The lowest value of access at inflection point is $\langle k\rangle_M=0.14$ for Marseille, and the largest $\langle k\rangle_M=0.79$ for Copenhagen ($\langle k\rangle_{M}=0.75$ for Oslo). However, for Oslo we have seen in Section \ref{sec:amsterdam} that the peak of average degree derivative is reached late in the interval, confirmed by the value of $\delta t_M=0.64$. Therefore, the value of $\langle k\rangle_M$ alone is not a sufficient indicator of access. These results indicate that for a more complete view of accessibility, a combination of primal and dual access indicators is necessary for assessing the overall network performance with respect to access.
For the 30-minute access indicators, the average degree at $t_b=30$ min ranges from $\langle k\rangle_{30}=0.11$ for New York to $\langle k\rangle_{30}=1.0$ for Genoa and Rennes. Since the maximum travel time varies largely across networks, the value of $\langle k\rangle_{30}=0.11$ is expected to be lower for cities with large values of $t_{max}$ (152 min for New York), and in the case of Genoa and Rennes the maximum travel time is below 30 minutes, with $t_{max}=26$ min and $t_{max}=20$ min, respectively.

\subsubsection*{JSD-based access indicators}

The biggest structural transition of the access graph, as reflected in the Jensen-Shannon divergence, consistently occurs earlier in the interval than the access inflection point. The values of the access structural change point $t_{JS}$ range from 3 minutes (Rennes) to 48 minutes (San Francisco). The normalised values range from $\delta t_{JS}=0.08$ for Bilbao and Valencia to $\delta t_{JS}=0.38$ for Baltimore. The values of access at structural change point range from $\langle k\rangle_{JS}=0.02$ for Bilbao to $\langle k\rangle_{JS}=0.35$ for San Francisco. An exception is the Genoa metro where $\langle k\rangle_{JS}=0.0$, due to the short time interval and simple degree change behaviour for a network with one line and 8 nodes. These results indicate that the largest structural reorganisation of the access graph, and thus the change in accessibility levels, will occur relatively early in the interval and consequently at low average degrees. In comparison, the point at the maximum growth of average degree often occurs when the structural changes are relatively small (e.g. Boston, Cairo, Oslo, or Washington, Figure \ref{fig:all_cities_indicators}). We also stress that whereas the average degree derivative vs. $t_b$ exhibits one distinctive peak for most cities (the exceptions are mostly the one line networks), the JSD shape is not as universal across networks and often exhibits another peak, or several peaks, later in the interval, which is not captured in the devised indicators (e.g. Oslo, Santiago, Stockholm).

\subsubsection*{Access equity indicators}

Next, we turn to investigating access equity. Values of inflection
access inequality $G_M$ and 30-min access inequality $G_{30}$ are in the interval between $\approx 0.05$ and $\approx 0.4$, with the exception of networks with $t_{max}\lesssim 30$ min where $G_{30}=0$. This happens for two of the one-line networks, Rennes and Genoa, as well as Copenhagen. The value of $G_M=0.14$ is a further indicator of the quality of the Copenhagen metro. The highest disparities are observed for Valencia ($G_M=0.40, G_{30}=0.33$) and Madrid ($G_M=0.37, G_{30}=0.40$). The largest value of $G_{30}=0.42$ is observed for New York however, with its large size and maximum travel time, this is expected and the value $G_M=0.25$ indicates moderate inequality.

\medskip

For the illustrative example of the Oslo and Vienna metros in the previous section (\ref{sec:amsterdam}), the values of the accessibility indicators are (for each indicator, the values are given first for Oslo, second for Vienna): $t_M=27$ and $t_M=59$; $\delta t_M=0.36$ and $\delta t_M=0.64$; $\langle k\rangle_{M}=0.46$ and $\langle k\rangle_{M}=0.75$; $\langle k\rangle_{30}=0.53$ and $\langle k\rangle_{30}=0.22$; $G_{M}=0.25$ and $G_{M}=0.15$; $G_{30}=0.23$ and $G_{30}=0.30$, confirming the qualitative observation of Vienna having higher overall accessibility as well as access equity than Oslo.

\subsection{Statistical analysis}

\begin{figure}[H]
	\begin{center}
		\texttt{}\includegraphics[scale=0.65]{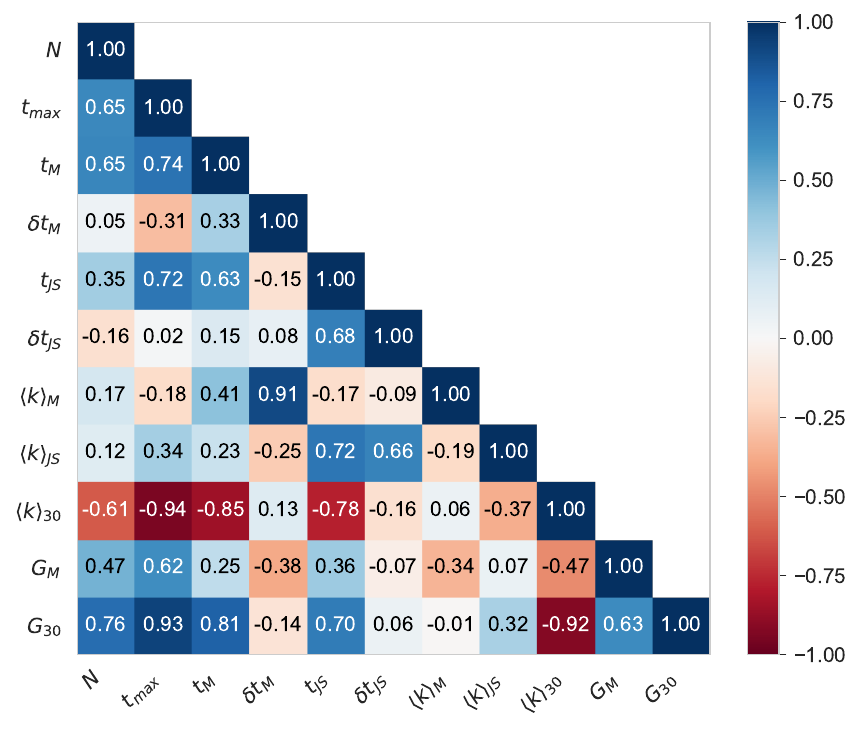}
	\end{center}
	\caption{Spearman correlation matrix for accessibility variables. $N$: number of nodes in the network; $t_{max}$: maximum generalised travel time (in minutes); $t_M$: access inflection point; $\delta t_M = \frac{t_M}{t_{max}}$ normalised access inflection point; $t_{JS}$: access structural change point; $\delta t_M = \frac{t_{JS}}{t_{max}}$ normalised access structural point; $\langle k\rangle_M$: access at inflection point; $\langle k\rangle_{JS}$: access at structural change point; $\langle k\rangle_{30}$: 30-min access; $G_M$: inflection access inequality; $G_{30}$: 30-min access inequality.
    }
	\label{fig:corr}
\end{figure}

\begin{figure}[H]
	\begin{center}
		\texttt{}\includegraphics[scale=0.5]{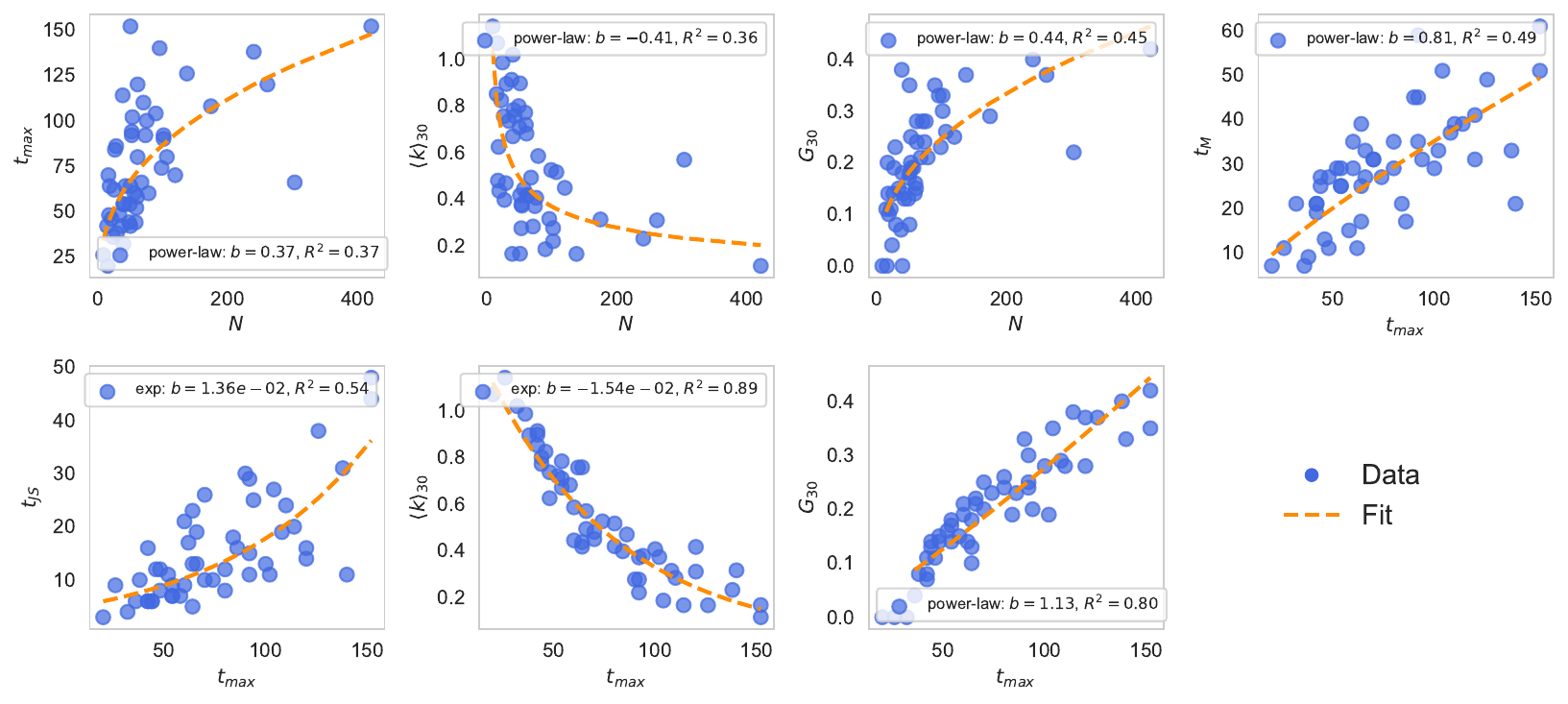}
	\end{center}
	\caption{Scatter plots for selected pairs of variables with the highest observed values of Spearman correlation coefficients. In each subplot, one data point corresponds to a metro network. Included are the best power-law or exponential fits with parameter values and goodness of fit. The fits are performed in log-log (log) scale for power-law (exponential) fits, and $R^2$ is determined in linear scale.
    }
	\label{fig:scatter}
\end{figure}

To better understand the relationships between different indicators, the Spearman correlation matrix for all variables is shown in Figure \ref{fig:corr}. Spearman correlation is chosen over Pearson due to non-linear relations between several pairs of variables. To examine in more detail the relations of the most significantly correlated pairs of variables with non-linear relationships, their scatter plots alongside the best power-law or exponential fits are shown in Figure \ref{fig:scatter}.

The maximum travel time $t_{max}$ varies per city and increases sub-linearly with the size of the network $N$ (Figure \ref{fig:scatter}). 
This indicates a relatively better performance for larger networks. This is expected due to the approximately universal values of individual travel time budgets, and in PTN planning $t_{max}$ must have a reasonable upper bound regardless of the size of the network. Note here that $N$ is correlated with city geographical area and population size.

Interestingly, the correlation of the access inflection point $t_M$ with its normalised counterpart $\delta t = t_M/t_{max}$ is relatively low ($r_S \approx 0.4$), suggesting the potential added value in using the combination of these indicators for measuring and comparing accessibility. Notable is the lack of correlation between $\delta t_M$ and $N$, indicating that this indicator offers a size-independent measure of access.

The correlations of $t_{JS}$ with $N$ is $r=0.35$, significantly lower than for $t_{M}$ with $N$. This suggests that the position of the time interval where the structural reorganisation of the access graph is the largest is largely independent of network size. In contrast, the correlation with $t_{max}$, $r=0.72$ is significant and comparable to $r=0.74$ for $t_M$ and $N$. The correlations of the normalised values $\delta t_{JS}$ with $N$ ($r=-0.16$) and $t_{max}$ ($r=0.02)$ are largely insignificant and indicate a lack of uniform behaviour of the JSD-based measures.

The values of the $30$-min type indicators, the 30-min access $\langle k\rangle_{30}$ 30-min access inequality $G_{30}$ correlate significantly with the size of the network $N$ and maximum generalised travel time $t_{max}$. The relations are shown in Figure \ref{fig:scatter}. Strong negative correlations and negative power-law relations are observed for $\langle k\rangle_{30}$ vs. $N$, while the shape of $\langle k\rangle_{30}$ vs. $t_{max}$ is best described with exponential decay. In contrast, the values of $G_{30}$ increase sub-linearly, following a power-law. For the largest networks, with size $N \gtrapprox 150$, the value of the Gini coefficient is approximately constant, indicating that beyond a certain size, larger networks tend to offer equally unequal distributions of access, in contrast to the relation observed for small and moderately sized networks (i.e. the growth of $G_M$ stalls after this value of $N$).

The strong negative correlation of $\langle k\rangle_{30}$ and $G_{30}$ complement the above observations, 
and the high positive correlation is expected due to similar patterns of behaviour for all networks.
For the primal indicators, the strong positive correlation of $\delta t$ with access at inflection point $\langle k\rangle_M$ points to a universal behaviour of average degree growth (observed to be logistic-like in most cases) across network sizes. The moderate positive correlation of inflection access inequality $G_M$, and $N$, and $G_M$ and $t_{max}$ indicate that network size does impact access inequality with larger networks tending to have larger disparities in access. Significant negative correlation between the Gini index and $\delta t$ indicates that inequity decreases with $\delta t$, complementing the previous observation on the relation to $N$.

\section{Discussion and conclusion}
\label{sec:conclusion}

We proposed a novel graph representation of public transport networks, the access graph $\mathcal{G}_A$, also denoted here as A-space. The edges in $\mathcal{G}_A$ are based on shortest paths weighted with generalised travel times, as determined from the weighted L- and P-space representations, and directly connect nodes, reachable from one another within a given time budget. We studied the time budget-dependent topology of the access graph, specifically node degree evolution. Average degree and degree distributions of the access graphs for 51 metro networks were examined and a set of global accessibility indicators was proposed. First, the time budget at the fastest growth of average degree $t_M$, and average degree at this time, $\langle k\rangle_M$, along with the relative value of the maximal growth time to maximum travel time, $\delta t_M$, examine the access graph behaviour at a critical point. Although the observed values are averaged quantities, the determination of the point at which they are observed follows from the global evolution of the whole access graph. The Gini coefficient $G_M$ of the degree distribution at the same value of time budget was used as an access equity indicator. These measures reflect network properties and can also be understood as network quality indicators. On the other hand, the value of the average degree $\langle k\rangle_{30}$ and the Gini coefficient $G_{30}$ at a fixed time budget of $30$ minutes offer a complementary view from the passengers' perspective, offering a yardstick for comparing networks. In addition to studying the averaged values, we examine the overall shape of the degree distribution and its change between consecutive time steps by observing the Jensen-Shannon divergence of the degree distribution. An important insight is that it offers an alternative perspective to the average degree change which amounts to a complementary metric.

The introduced approach addresses an important gap in accessibility research. Our operationalisation of accessibility, based on graph theory and network science, contributes a methodology to unify accessibility studies. Instead of devising intricate metrics to include different properties of the transport system in the existing representations, we propose an algorithmic approach where each step is well-defined, and the end result is a new graph representation where the simplest metric, the node degree, offers reliable, interpretable, and transferable indicators. The widespread availability of standardised GTFS data is leveraged to build the standard PT graph representations, on which our methodology is based. This makes the proposed approach readily generalisable to other networks of all PT modalities at different scales. Importantly, the methodology can be applied with modifications of each step, for example, using different edge weights or different functions of distance, and arrive at similar definitions of indicators without compromising interpretability. Moreover, the observed similarity in behaviour of the proposed measures across all 51 metro networks included in our analysis, supports the relevance of our approach for accessibility analysis.

Alongside the aforementioned contributions of this work, our study is also subject to certain limitations which we discuss in the following and outline several related venues for future research to address those. The most obvious omission of the model is the exclusion of data on the number of opportunities for different activities available at accessible destinations. The access graph in this original form addresses only the transport component of accessibility, implicitly assuming the number of reachable stops as an approximation for opportunities of activity. Including land use data in the model and developing suitable indicators for this extended representation constitutes an important important further research direction.

Furthermore, the geographical scale of the networks can jeopardise the comparison of networks of such varying sizes as in our data set, especially for absolute-value or fixed-time indicators, such as $t_{max}$, $t_M$ or $\langle k\rangle_{30}$. Notwithstanding our effort to address this issue to a certain extent by introducing the normalised versions of average degree and the times relative to the maximum time, this leaves open the normative question of the maximum generalised travel time as an accessibility indicator in itself (i.e. does the observed $t_{max}$ make sense for the geographical scope of this network?). This has been examined in detail in a follow-up study \cite{vsfiligoj2026network}, where we compare the access behaviour of each metro network to its idealised counterpart based on maximally connected subgraphs of the access graph.

Finally, the access graph is modelled as an unweighted undirected graph. This is the result of taking the simplest decay function of distance, i.e. the cut-off function. In reality, individual time budgets and weights for different components of travel time may vary across passengers. Taking a probabilistic approach to edge inclusion or using fuzzy cut-off values will be an interesting next step. 







We see several promising generalisations in the direction of multilayer networks. The structure of the access graph itself hints at multilayer networks; we believe there is large potential in bridging the theory of multilayer networks and accessibility to address the problem in its full complexity. First, the formalisation of the access graph as a temporal network will enable ready application of existing methods and deeper insights, such as temporal motif detection \cite{paranjape2017motifs}.
Second, the access graph can be used to model different types of interactions, such as financial costs, perceived accessibility, etc. Including layers with different thresholds for different variables will be an interesting generalisation, such as modelling the X-minute, Y-cost city \cite{chen2025evaluating}. Heterogeneous networks where different entities (stops, routes, passengers, etc.) can be modelled to interact with each other can guide the development of accessibility studies. 

In a wider sense, the access graph can be seen as introducing a method to formulate graph representations with functional connections which was recently reviewed for robustness-related studies in \citep{buth2025functional}. Modelling the PT system properties at the level of network structure, not at the level of its properties, itself emerges as a promising direction for PTN studies.

We note that the relationships between the introduced access indicators and the size of the network, as reflected in its topological dimension $N$ and network diameter $t_{max}$ observed in Figure \ref{fig:scatter} exhibit power-law behaviour in most cases. Although the fits do not exhibit strong statistical significance, they point to (weak) evidence of allometric scaling \citep{samaniego2008cities}. The analysis of access behaviour with the focus on maximal achievable levels of access within the framework of allometric scaling offers an interesting venue of future research. Beyond (public) transport networks, we note the potential of the Access graph methodology to inspire notions similar to accessibility and network science approaches in other application domains, complementing related approaches such as that in \citep{lentz2013unfolding}.

The new representation can also guide planners and policy makers in designing accessible public transport networks. First, by transferring the access formalism into the network structure, the indicators it offers are highly interpretable and reproducible across all modes of PT systems. The L- and P-space representations, on which the access graph construction is based, use the GTFS data, making the construction of the access graph readily available. Detailed analysis and visualisation can help identify regions with lower access levels, and simulations can provide strategies for equitable planning in the strategic or tactical stage by studying the impact of changes to the network.




\section*{Acknowledgements}
The first and second authors were partially supported by ARIS-Slovenian Agency for Research and Innovation (grants P2-0394 (A) and P1-0222, respectively, and grant SN-ZRD/22-27/0510 for both authors).

\appendix






\section{Sensitivity analysis for JSD}
\label{app:sens}

\begin{figure}[H]
	\begin{center}
		\texttt{}\includegraphics[scale=0.55]{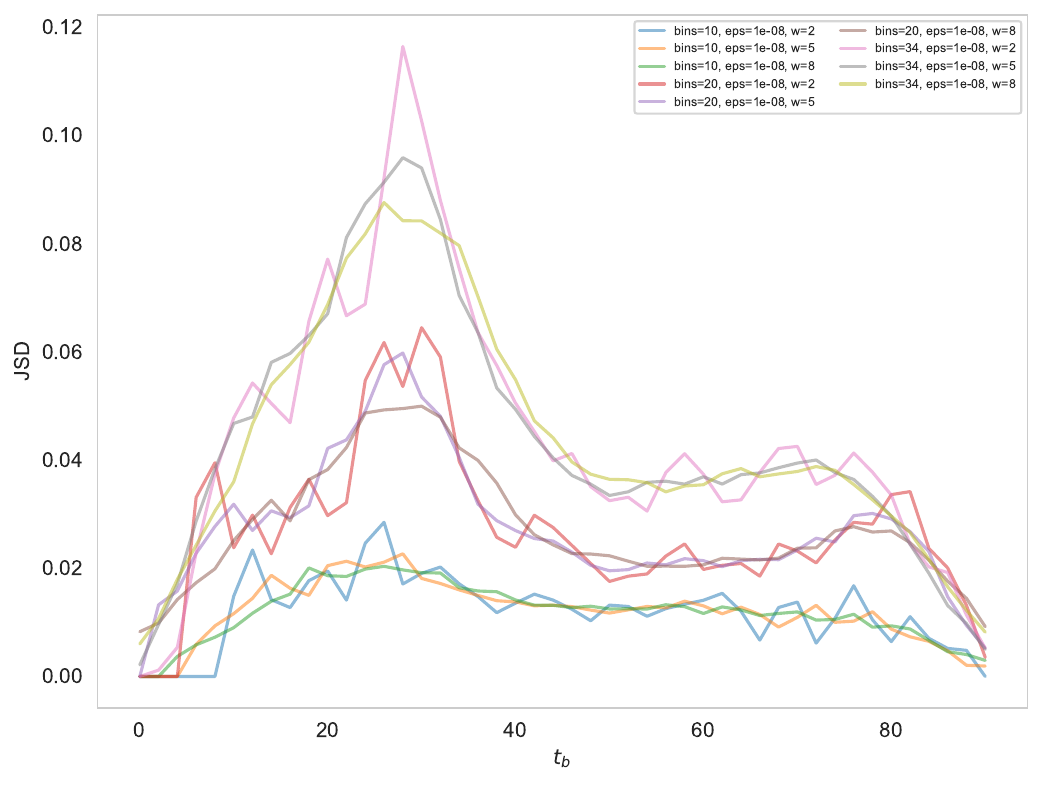}
	\end{center}
	\caption{JSD sensitivity analysis for the Oslo network; sliding time window and binning are varied.}
	\label{fig:sensoslo}
\end{figure}

\begin{figure}[H]
	\begin{center}
		\texttt{}\includegraphics[scale=0.6]{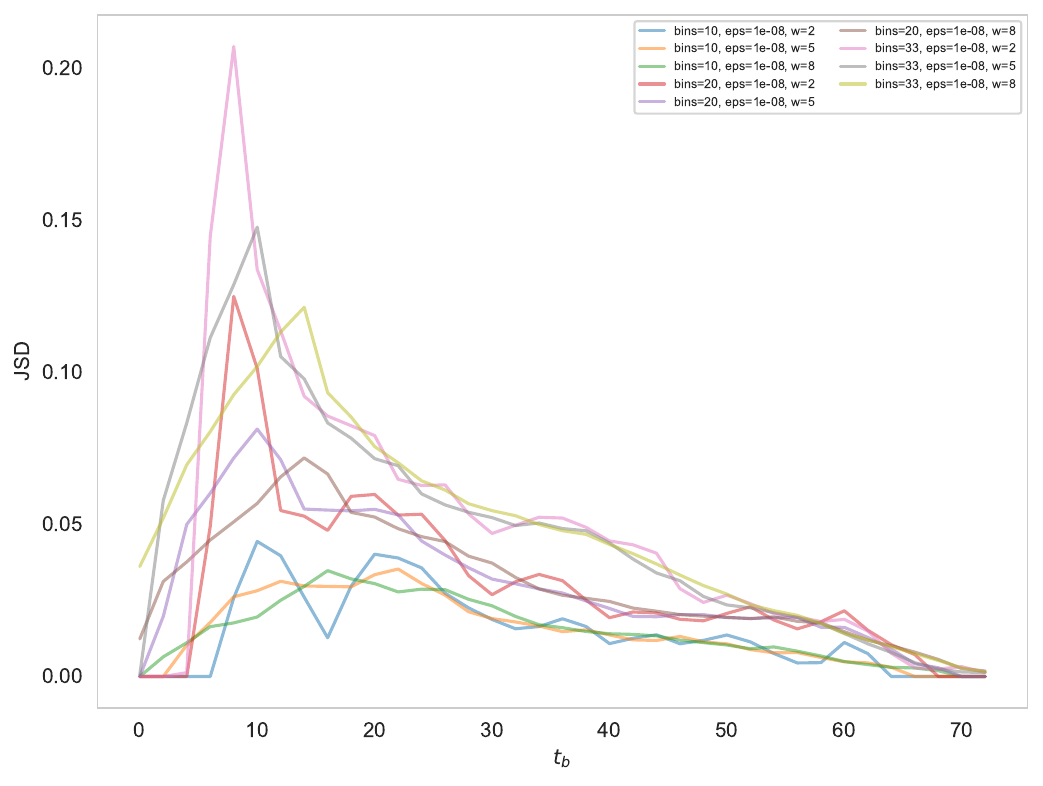}
	\end{center}
	\caption{JSD sensitivity analysis for the Vienna network; sliding time window and binning are varied.}
	\label{fig:sensvienna}
\end{figure}

Here, we provide sensitivity analysis for the JSD analysis for the Oslo and Vienna networks. For our results, we used a fixed binning with $n_{bins} = \text{round}(N/5)$ to construct the degree distribution and a sliding window of width 5 to smoothen the curve $JSD(t_b)$. Here we systematically vary the binning and width of the sliding window. The results suggest robustness w.r.t. both parameters for both cities as the shapes and the location(s) of the peaks stay similar across all combinations of parameters (Figure \ref{fig:sensoslo} for Oslo and Figure \ref{fig:sensvienna} for Vienna). When the binning is too coarse, the shapes show less variance, but the binning of only 10 bins distorts the distributions. As the number of bins rises and the point-wise variations become larger the shapes remain similar in all cases.

\bibliography{references}

\end{document}